\documentclass[aps,superscriptaddress]{revtex4}
\usepackage[utf8]{inputenc}
\usepackage[a4paper]{geometry}
\usepackage{float}
\usepackage{amsmath}
\usepackage{amssymb}
\usepackage{graphicx}
\usepackage{esint}

\makeatletter

\providecommand{\tabularnewline}{\\}

\@ifundefined{textcolor}{}
{%
 \definecolor{BLACK}{gray}{0}
 \definecolor{WHITE}{gray}{1}
 \definecolor{RED}{rgb}{1,0,0}
 \definecolor{GREEN}{rgb}{0,1,0}
 \definecolor{BLUE}{rgb}{0,0,1}
 \definecolor{CYAN}{cmyk}{1,0,0,0}
 \definecolor{MAGENTA}{cmyk}{0,1,0,0}
 \definecolor{YELLOW}{cmyk}{0,0,1,0}
}

\providecommand{\tabularnewline}{\\}

\makeatother

\begin{document}

\title{Approach to Chandrasekhar-Kendall-Woltjer State in a Chiral Plasma}

\author{Xiao-liang Xia}

\affiliation{Department of Modern Physics, University of Science and Technology
of China, Hefei, Anhui 230026, China}

\author{Hong Qin}

\affiliation{Department of Modern Physics, University of Science and Technology
of China, Hefei, Anhui 230026, China}

\affiliation{Plasma Physics Laboratory, Princeton University, P.O. Box 451, Princeton,
New Jersey 08543, USA}

\author{Qun Wang}

\affiliation{Department of Modern Physics, University of Science and Technology
of China, Hefei, Anhui 230026, China}
\begin{abstract}
We study the time evolution of the magnetic field in a plasma with
a chiral magnetic current. The Vector Spherical Harmonic functions
(VSH) are used to expand all fields. We define a measure for the Chandrasekhar-Kendall-Woltjer
(CKW) state, which has a simple form in VSH expansion. We propose
the conditions for a general class of initial momentum spectra that
will evolve into the CKW state. For this class of initial conditions,
to approach the CKW state, (i) a non-vanishing chiral magnetic conductivity
is necessary, and (ii) the time integration of the product of the
electric resistivity and chiral magnetic conductivity must grow faster
than the time integration of the resistivity. We give a few examples
to test these conditions numerically which work very well. 
\end{abstract}
\maketitle

\section{Introduction}

In high energy heavy-ion collisions, two heavy nuclei are accelerated
to almost the speed of light and produce very strong electric and
magnetic fields at the moment of the collision \cite{Kharzeev:2007jp,Skokov:2009qp,Voronyuk:2011jd,Deng:2012pc,Bloczynski:2012en,McLerran:2013hla,Gursoy:2014aka,Roy:2015coa,Tuchin:2014iua,Li:2016tel}.
The magnitude of magnetic fields can be estimated as $eB\sim\gamma vZe^{2}/R_{A}^{2}$,
where $Z$ and $R_{A}$ are the proton number and the radius of the
nucleus respectively, $v\approx1-2 m_{n}^{2}/s$ is the the velocity
of the nucleus and $\gamma$ is the Lorentz contraction factor $\gamma\approx\sqrt{s/2}/m_{n}$
($m_{n}$ is the nucleon mass and $\sqrt{s}$ is the collision energy
per nucleon). In Au+Au collisions at the Relativistic Heavy Ion Collider
(RHIC) with $\sqrt{s}=200$ GeV, the peak value of the magnetic field
at the moment of the collision is about $5m_{\pi}^{2}$ ($m_{\pi}$
is the pion mass) or $1.4\times10^{18}$ Gauss. In Pb+Pb collisions
at the Large Hadron Collider (LHC) with $\sqrt{s}=2.76$ TeV, the
peak value of the magnetic field can be 10 times as large as at RHIC.
Such high magnetic fields enter strong interaction regime and may
have observable effects on the hadronic events. The chiral magnetic
effect (CME) is one of them which is the generation of an electric
current induced by magnetic fields from of the imbalance of chiral
fermions \cite{Kharzeev:2007jp,Fukushima:2008xe,Kharzeev:2015znc}.
The CME and other related effects have been widely studied in quark-gluon
plasma produced in heavy-ion collisions. The charge separation effect
observed in STAR \cite{Abelev:2009ac,Abelev:2009ad} and ALICE \cite{Abelev:2012pa}
experiments are consistent to the CME predictions, although there
may be other sources such as collective flows that contribute to the
charge separation \cite{Huang:2015fqj}. The CME has recently been
confirmed to exist in materials such as Dirac and Weyl semi-metals
\cite{Son:2012bg,Basar:2013iaa,Li:2014bha}.

In hot and dense matter an imbalance in the number of right-handed
quarks and left-handed quarks may be produced through transitions
between vacua of different Chern-Simons numbers in some domains of
the matter. This is called chiral anomaly and is described by the
anomalous conservation law for the axial current, 
\begin{equation}
\partial_{\mu}j_{A}^{\mu}=-\frac{N_{c}e^{2}\sum_{f}Q_{f}^{2}}{8\pi^{2}}F_{\mu\nu}\tilde{F}^{\mu\nu}-\frac{N_{f}g^{2}}{16\pi^{2}}F_{\mu\nu}^{a}\tilde{F}_{a}^{\mu\nu},\label{eq:anomaly}
\end{equation}
where $j_{A}^{\mu}=(n_{A},\mathbf{j}_{A})$ denotes the axial 4-vector
current with $n_{A}$ being the chiral charge, $N_{c}$ and $N_{f}$
are the number of colors and flavors of quarks respectively, $Q_{f}$
is the electric charge (in the unit of electron charge $e$) of the
quark with flavor $f$, $F_{\mu\nu}$ denotes the field strength of
the electromagnetic field and $\tilde{F}_{\mu\nu}=\frac{1}{2}\epsilon^{\mu\nu\rho\sigma}F_{\rho\sigma}$
is its dual, $g$ is the strong coupling constant, $F_{\mu\nu}^{a}$
denotes the field strength of the $a$-th gluon with $a=1,2,\cdots,N_{c}^{2}-1$
and $\tilde{F}_{a}^{\mu\nu}=\frac{1}{2}\epsilon^{\mu\nu\rho\sigma}F_{\rho\sigma}^{a}$
is its dual. The first term on the right-hand-side of Eq. (\ref{eq:anomaly})
is the anomaly term from electromagnetic fields while the second one
is from gluonic fields. In Eq. (\ref{eq:anomaly}) we have neglected
quark masses. For electromagnetic fields we can write $F_{\mu\nu}\tilde{F}^{\mu\nu}$
in the 3-vector form using $F_{\mu\nu}\tilde{F}^{\mu\nu}=-4\mathbf{E}\cdot\mathbf{B}$,
where $\mathbf{E}$ and $\mathbf{B}$ are the electric and magnetic
3-vector field respectively.

The axial current $j_{A}^{\mu}$ breaks the parity locally and may
appear in one event, but it is vanishing when taking event average.
With such an imbalance, an electric current can be induced along the
magnetic field, so the total electric current can be written as 
\begin{equation}
\mathbf{j}=\sigma\mathbf{E}+\sigma_{\chi}\mathbf{B},\label{eq:current}
\end{equation}
where $\sigma$ and $\sigma_{\chi}$ are the electric and chiral conductivity
respectively. Note that $\sigma_{\chi}$ is proportional to the difference
between the number of right-handed quarks and left-handed quarks which
breaks the parity but conserves the time reversal symmetry. This is
in contrast with the electric conductivity which breaks the time reversal
symmetry but conserves the parity. So the Ohm's current is dissipative
(with heat production) while the chiral magnetic current is non-dissipative.

We consider a system of charged fermions in electromagnetic fields.
The term of $F_{\mu\nu}\tilde{F}^{\mu\nu}$ or $\mathbf{E}\cdot\mathbf{B}$
in Eq. (\ref{eq:anomaly}) is actually related to the magnetic helicity
$H_{\text{mag}}=\int d^{3}x\mathbf{A}\cdot\mathbf{B}$. Then we can
take volume integration of Eq. (\ref{eq:anomaly}) and obtain 
\begin{equation}
\frac{d}{dt}H_{\text{total}}=0,
\end{equation}
where the total helicity $H_{\text{total}}$ is defined by combining
the magnetic helicity and the chiral charge $N_{A}=\int d^{3}x\, n_{A}$,
\begin{equation}
H_{\text{total}}=H_{\text{mag}}-C_{A}^{-1}N_{A},\label{eq:conservation}
\end{equation}
where $C_{A}\equiv\frac{N_{c}e^{2}\sum_{f}Q_{f}^{2}}{2\pi^{2}}$.
This means that the magnetic helicity and the chiral charge of fermions
can be transferred into each other.

The Chandrasekhar-Kendall-Woltjer (CKW) state is a state of the magnetic
field which satisfies the following equation 
\begin{equation}
\nabla\times\mathbf{B}=C\mathbf{B},\label{eq:ckw}
\end{equation}
where $C$ is a constant. The CKW state was first studied by Chandrasekhar,
Kendall and Woltjer \cite{chandrasekhar1956force,chandrasekhar1957force,chandrasekhar1958force,woltjer1958theorem}
as a force free state. We notice that in a plasma with the chiral
magnetic current (\ref{eq:current}), if the Ohm's current is absent,
the system reaches a special CKW state with $C=\sigma_{\chi}$ following
the Ampere's law. To our knowledge, this idea was first proposed in
\cite{Chernodub:2010ye}. But with the Ohm's current, can the CKW
state still be reached? This question can be re-phrased as: what are
the conditions under which the CKW state can be reached in a plasma
with chiral magnetic currents? In this paper we will answer this question
by studying the evolution of magnetic fields with the Maxwell-Chern-Simons
equations.

In classical plasma physics, a state satisfying Eq. (\ref{eq:ckw})
is called the Taylor state or the Woltjer-Taylor state. It was first
found by Woltjer \cite{woltjer1958theorem} that the CKW state minimizes
the magnetic energy for a fixed magnetic helicity. In toroidal plasma
devices, such a state is often observed as a self-generated state
called reverse field pinch with the distinct feature that the toroidal
fields in the center and the edge point to opposite directions. Taylor
\cite{Taylor74,Taylor86} first argued that the minimization of magnetic
energy with a fix magnetic helicity is realized as a selective decay
process in a weakly dissipative plasma when the dynamics is dominated
by short wavelength structures. Taylor's theory has been questioned
and debated intensely, and alternative theories has been proposed
\cite{Ortolani93-56,Qin12PRL}. It is certainly interesting that both
classical plasmas and chiral plasmas have the tendency to evolve towards
such a state, which suggests that the two systems may share certain
dynamics features responsible for the emerging of the state. A brief
discussion in this aspect is given in the paper as well.

The paper is organized as follows. In Section \ref{sec:mcs_eq}, we
start with the Maxwell-Chern-Simons equations and define a global
measure for the CKW state by the magnetic field and the electric current.
In Section \ref{sec:vsh}, we expand all fields in Vector Spherical
Harmonic (VSH) functions. The inner products have simple forms in
the VSH expansion. The parity of a quantity can be easily identified
in the VSH form. We give in Section \ref{sec:mcs_vsh} the solution
to the Maxwell-Chern-Simons equations for each mode in the VSH expansion.
The conditions for the CKW state are given in Section \ref{sec:cond}.
In Section \ref{sec:examples} we test these conditions by examples
including the ones with constant and self-consistently determined
$\sigma_{\chi}$. We also generalize the momentum spectra at the initial
time from a power to a polynomial pre-factor of scalar momentum in
Section \ref{sec:polynomial}. The summary and conclusions are made
in the last section.

\section{Maxwell-Chern-Simons equations and CKW state}

\label{sec:mcs_eq}We start from Maxwell-Chern-Simons equations or
anomalous Maxwell equations, 
\begin{align}
\nabla\times\mathbf{B} & =\sigma\mathbf{E}+\sigma_{\chi}\mathbf{B},\label{eq:Maxwell_eq1}\\
\nabla\times\mathbf{E} & =-\frac{\partial}{\partial t}\mathbf{B},\label{eq:Maxwell_eq2}\\
\nabla\cdot\mathbf{B} & =0,\label{eq:Maxwell_eq3}\\
\nabla\cdot\mathbf{E} & =0,
\end{align}
where we have included the induced current $\mathbf{j}=\sigma\mathbf{E}+\sigma_{\chi}\mathbf{B}$
and neglected the displacement current $\frac{\partial}{\partial t}\mathbf{E}$.
We have also dropped the external charge and current density. We assume
that $\sigma$ and $\sigma_{\chi}$ depend on $t$ only.

Taking curl of Eq. (\ref{eq:Maxwell_eq1}) and using Eqs. (\ref{eq:Maxwell_eq2},\ref{eq:Maxwell_eq3}),
we obtain 
\begin{equation}
\sigma\frac{\partial}{\partial t}\mathbf{B}=\nabla^{2}\mathbf{B}+\sigma_{\chi}\nabla\times\mathbf{B}.\label{eq:short_eq}
\end{equation}
A similar equation for $\mathbf{E}$ can also be derived but we will
not consider it in the current study. To measure whether the CKW state
is reached in the evolution of the magnetic field, we introduce the
quantity 
\begin{equation}
\cos^{2}\theta(t)\equiv\frac{\left[\int d^{3}x(\nabla\times\mathbf{B})\cdot\mathbf{B}\right]^{2}}{\int d^{3}x(\nabla\times\mathbf{B})\cdot(\nabla\times\mathbf{B})\int d^{3}x\mathbf{B}\cdot\mathbf{B}}=\frac{\left\langle \mathbf{j},\mathbf{B}\right\rangle ^{2}}{\left\langle \mathbf{j},\mathbf{j}\right\rangle \left\langle \mathbf{B},\mathbf{B}\right\rangle },\label{eq:cos_def}
\end{equation}
where we have used the notation for the inner product $\left\langle \mathbf{X},\mathbf{Y}\right\rangle \equiv\int d^{3}x(\mathbf{X}\cdot\mathbf{Y})$
for any vector field $\mathbf{X}$ and $\mathbf{Y}$. According to
the Cauchy-Schwartz inequality 
\begin{equation}
\left\langle \mathbf{j},\mathbf{j}\right\rangle \left\langle \mathbf{B},\mathbf{B}\right\rangle \geq\left\langle \mathbf{j},\mathbf{B}\right\rangle ^{2},
\end{equation}
we have $\cos^{2}\theta(t)\leq1$, where the equality holds only in
the case of $\mathbf{j}\parallel(\pm\mathbf{B})$ when the CKW state
is reached \cite{Qin12PRL}. We assume that $\cos^{2}\theta(t)$ is
a smooth function of $t$. The condition that the CKW state is reached
can be given by 
\begin{equation}
\lim_{t\to\infty}\cos^{2}\theta(t)=1.
\end{equation}
Note that $\cos^{2}\theta(t)$ should not exactly be equal to 1, since
$\cos^{2}\theta(t)$ is a smooth function of $t$ and bounded by the
upper limit 1.

To see the time evolution of $\cos^{2}\theta(t)$, it is helpful to
write inner products in simple forms, which we will do in the next
section.

\section{Expansion in vector spherical harmonic functions}

\label{sec:vsh}In this section, we will expand all fields in the
basis of Vector Spherical Harmonic function (VSH), with which we can
put inner products into a simple and symmetric form.

\subsection{Expansion in VSH}

The quantities we use to express $\cos^{2}\theta(t)$ in Eq. (\ref{eq:cos_def})
are $\mathbf{B}=\nabla\times\mathbf{A}$ and $\mathbf{j}=\nabla\times\mathbf{B}=\nabla\times(\nabla\times\mathbf{A})$.
We can extend the series to include more curls, 
\begin{equation}
\mathbf{A},\ \mathbf{B}=\nabla\times\mathbf{A},\ \mathbf{j}=(\nabla\times)^{2}\mathbf{A},\cdots,\ (\nabla\times)^{n}\mathbf{A},\cdots,\label{eq:A_series}
\end{equation}
So the inner products can be written as $\left\langle (\nabla\times)^{n_{1}}\mathbf{A},(\nabla\times)^{n_{2}}\mathbf{A}\right\rangle $
with $n_{1}$ and $n_{2}$ are non-negative integers. To find an unified
form for the fields in this series, we can expand $\mathbf{A}(\mathbf{x},t)$
in the Coulomb gauge in VSH 
\begin{equation}
\mathbf{A}(\mathbf{x},t)=\frac{1}{\pi}\sum_{l,m}\int_{0}^{\infty}dk\, k\left[\alpha_{lm}^{+}(t,k)\mathbf{W}_{lm}^{+}(\mathbf{x};k)-\alpha_{lm}^{-}(t,k)\mathbf{W}_{lm}^{-}(\mathbf{x};k)\right],\label{eq:A_expand}
\end{equation}
where $k=|\mathbf{k}|$ is the scalar momentum and $l,m$ are the
quantum number of the angular momentum and the angular momentum along
a particular direction respectively. $\mathbf{W}_{lm}^{\pm}(\mathbf{x};k)$
are divergence-free vector fields which can be expressed in term of
VSH $\mathbf{X}_{lm}=\frac{-i}{\sqrt{l(l+1)}}\mathbf{r}\times\nabla Y_{lm}$
\cite{jackson1999classical}. The explicit form of $\mathbf{W}_{lm}^{\pm}$
can be found in, e.g., Ref. \cite{Hirono:2015rla,Tuchin:2016tks}.
The orthogonal basis functions $\mathbf{W}_{lm}^{\pm}(\mathbf{x};k)$
satisfy the following orthogonality relations, 
\begin{equation}
\int d^{3}x\mathbf{W}_{l_{1}m_{1}}^{s_{1}*}(\mathbf{x};k)\cdot\mathbf{W}_{l_{2}m_{2}}^{s_{2}}(\mathbf{x};k^{\prime})=\frac{\pi}{k^{2}}\delta(k-k^{\prime})\delta_{l_{1}l_{2}}\delta_{m_{1}m_{2}}\delta_{s_{1}s_{2}},\label{eq:w_orth}
\end{equation}
where $s_{1},s_{2}=\pm$. Note that $\mathbf{W}_{lm}^{\pm}(\mathbf{x};k)$
themselves are CKW states satisfying 
\begin{equation}
\nabla\times\mathbf{W}_{lm}^{\pm}(\mathbf{x};k)=\pm k\mathbf{W}_{lm}^{\pm}(\mathbf{x};k),\label{eq:w_curl}
\end{equation}
and are divergence-free, $\nabla\cdot\mathbf{W}_{lm}^{\pm}=0$, so
we can expand any divergence-free vector fields in $\mathbf{W}_{lm}^{\pm}$.
We note that $\mathbf{A}(\mathbf{x},t)$ is real while $\alpha_{lm}^{\pm}(t,k)$
and $\mathbf{W}_{lm}^{\pm}(\mathbf{x};k)$ are complex.

Taking curls of Eq. (\ref{eq:A_expand}) and using Eq. (\ref{eq:w_curl}),
we can expand all quantities in the series (\ref{eq:A_series}) in
VSH, 
\begin{equation}
(\nabla\times)^{n}\mathbf{A}=\frac{1}{\pi}\sum_{l,m}\int_{0}^{\infty}dk\, k^{n+1}\left[\alpha_{lm}^{+}(t,k)\mathbf{W}_{lm}^{+}(\mathbf{x};k)+(-1)^{n+1}\alpha_{lm}^{-}(t,k)\mathbf{W}_{lm}^{-}(\mathbf{x};k)\right].
\end{equation}
In particular the expansions of $\mathbf{B}$ and $\mathbf{j}$ are
\begin{align}
\mathbf{B}(\mathbf{x},t) & =\frac{1}{\pi}\sum_{l,m}\int_{0}^{\infty}dk\, k^{2}\left[\alpha_{lm}^{+}(t,k)\mathbf{W}_{lm}^{+}(\mathbf{x};k)+\alpha_{lm}^{-}(t,k)\mathbf{W}_{lm}^{-}(\mathbf{x};k)\right],\nonumber \\
\mathbf{j}(\mathbf{x},t) & =\frac{1}{\pi}\sum_{l,m}\int_{0}^{\infty}dk\, k^{3}\left[\alpha_{lm}^{+}(t,k)\mathbf{W}_{lm}^{+}(\mathbf{x};k)-\alpha_{lm}^{-}(t,k)\mathbf{W}_{lm}^{-}(\mathbf{x};k)\right].\label{eq:Bj_expand}
\end{align}
Note that the signs in front of $\alpha_{lm}^{-}$ are $(-1)^{n+1}$,
which are related to the parity of the quantity.

\subsection{Inner products}

We can put the general inner products into a simple form by using
the orthogonality relation (\ref{eq:w_orth}), 
\begin{equation}
\left\langle (\nabla\times)^{n_{1}}\mathbf{A},(\nabla\times)^{n_{2}}\mathbf{A}\right\rangle =\int_{0}^{\infty}dk\, k^{n_{1}+n_{2}}\left[g_{+}(t,k)+(-1)^{n_{1}+n_{2}}g_{-}(t,k)\right],\label{eq:inner_product_integral}
\end{equation}
where $g_{\pm}(t,k)$ are defined by 
\begin{equation}
g_{\pm}(t,k)=\frac{1}{\pi}\sum_{l,m}|\alpha_{lm}^{\pm}(t,k)|^{2}.
\end{equation}
We note that $g_{\pm}(t,k)$ are positive definite. In deriving Eq.
(\ref{eq:inner_product_integral}) we have used the fact that $\mathbf{A}$
is real so that $\left\langle (\nabla\times)^{n_{1}}\mathbf{A},(\nabla\times)^{n_{2}}\mathbf{A}\right\rangle =\left\langle (\nabla\times)^{n_{1}}\mathbf{A}^{*},(\nabla\times)^{n_{2}}\mathbf{A}\right\rangle $.
For convenience, we use the following short-hand notation for the
integral 
\begin{equation}
\int k^{n}\equiv\int_{0}^{\infty}dk\, k^{n}\left[g_{+}(t,k)+(-1)^{n}g_{-}(t,k)\right].
\end{equation}
From Eq. (\ref{eq:inner_product_integral}) it is easy to verify 
\begin{equation}
\left\langle (\nabla\times)^{n_{1}}\mathbf{A},(\nabla\times)^{n_{2}}\mathbf{A}\right\rangle =\left\langle (\nabla\times)^{n_{3}}\mathbf{A},(\nabla\times)^{n_{4}}\mathbf{A}\right\rangle 
\end{equation}
for $n_{1}+n_{2}=n_{3}+n_{4}$. This is consistent to the identity
$\left\langle \nabla\times\mathbf{X},\mathbf{Y}\right\rangle =\left\langle \mathbf{X},\nabla\times\mathbf{Y}\right\rangle $.
In Table (\ref{tab:inner-products}), we list the VSH forms of some
inner products that we are going to study later in this paper.

\begin{table}[H]
\centering{}%
\begin{tabular}{|c|c|c|c|}
\hline 
Quantity  & VSH form  & Short-hand  & Parity\tabularnewline
\hline 
$\left\langle \mathbf{j},\mathbf{j}\right\rangle $  & $\int_{0}^{\infty}dk\, k^{4}\left[g_{+}(t,k)+g_{-}(t,k)\right]$  & $\int k^{4}$  & even\tabularnewline
\hline 
$\left\langle \mathbf{j},\mathbf{B}\right\rangle $  & $\int_{0}^{\infty}dk\, k^{3}\left[g_{+}(t,k)-g_{-}(t,k)\right]$  & $\int k^{3}$  & odd\tabularnewline
\hline 
$\left\langle \mathbf{B},\mathbf{B}\right\rangle $  & $\int_{0}^{\infty}dk\, k^{2}\left[g_{+}(t,k)+g_{-}(t,k)\right]$  & $\int k^{2}$  & even\tabularnewline
\hline 
$\left\langle \mathbf{A},\mathbf{B}\right\rangle $  & $\int_{0}^{\infty}dk\, k\left[g_{+}(t,k)-g_{-}(t,k)\right]$  & $\int k^{1}$  & odd\tabularnewline
\hline 
\end{tabular}\caption{\label{tab:inner-products}Some inner products in VSH expansion. }
\end{table}

\subsection{Parity and helicity}

From Eq. (\ref{eq:w_curl}), the parity transformation $\mathbf{x}\to-\mathbf{x}$
is equivalent to the interchange of the $+$ and $-$ mode. In the
series (\ref{eq:A_series}), the quantity $(\nabla\times)^{n}\mathbf{A}$
is parity-even/parity-odd (P-even/P-odd) for odd/even $n$. For instance,
$\mathbf{A}$, $\mathbf{B}$ and $\mathbf{j}$ are P-odd, P-even and
P-odd respectively.

Also the inner product $\int k^{n}$ is P-even/P-odd for even/odd
$n$, see the last column of Table (\ref{tab:inner-products}) where
the magnetic helicity $H=\left\langle \mathbf{A},\mathbf{B}\right\rangle =\int k$
is P-odd, and the magnetic energy $W=\frac{1}{2}\left\langle \mathbf{B},\mathbf{B}\right\rangle =\frac{1}{2}\int k^{2}$
is P-even.

For a momentum spectrum containing only $g_{+}$, the helicity is
positive. If such a magnetic field can approach the CKW state, it
means $\mathbf{j}\parallel\mathbf{B}$ because $\left\langle \mathbf{j},\mathbf{B}\right\rangle $
is also positive for $g_{+}$ mode. In contrast it would mean $\mathbf{j}\parallel-\mathbf{B}$
for a momentum spectrum containing only $g_{-}$.

\section{Solving Maxwell-Chern-Simons equations in VSH}

\label{sec:mcs_vsh}The evolution equation of $\mathbf{B}$ in (\ref{eq:short_eq})
can be transformed into equations for $\alpha_{lm}^{\pm}(t,k)$ or
equivalently $g_{\pm}(t,k)$ with Eqs. (\ref{eq:Bj_expand},\ref{eq:w_orth}),
\begin{align}
\frac{d}{dt}\alpha_{lm}^{\pm}(t,k) & =\eta(-k^{2}\pm\sigma_{\chi}k)\alpha_{lm}^{\pm}(t,k),\nonumber \\
\frac{d}{dt}g_{\pm}(t,k) & =2\eta(-k^{2}\pm\sigma_{\chi}k)g_{\pm}(t,k),\label{eq:d_g/dt}
\end{align}
where $\eta=\frac{1}{\sigma}$ is the electric resistivity.

The solution of $g_{\pm}(t,k)$ is in the form 
\begin{equation}
g_{\pm}(t,k)=g_{\pm}(t_{0},k)e^{-2k^{2}\Lambda(t)\pm2k\Theta(t)},\label{eq:g_behavior}
\end{equation}
where $g_{\pm}(t_{0},k)$ denote the values at the initial time $t_{0}$,
and $\Lambda(t)$ and $\Theta(t)$ are defined by 
\begin{equation}
\Lambda(t)=\int_{t_{0}}^{t}\eta(\tau)d\tau,\quad\Theta(t)=\int_{t_{0}}^{t}\eta(\tau)\sigma_{\chi}(\tau)d\tau.\label{eq:lambda_theta}
\end{equation}
Note that both $\Lambda(t)$ and $\Theta(t)$ are positive.

Alternatively we can rescale time by using $\Lambda$ as a new evolution
parameter, and rewrite $\Theta(\Lambda)=\int_{0}^{\Lambda}\sigma_{\chi}(\Lambda)d\Lambda$,
i.e., $\Theta(\Lambda)$ is the integrated value of $\sigma_{\chi}(t)$
from $0$ to $\Lambda$.

There is a competition between $\sigma$ and $\sigma_{\chi}$ for
approaching or departing the CKW state. Large values of $\Theta(t)$
is favored for the CKW state. We will show that it is indeed determined
by the increasing ratio of $\Lambda(t)$ to $\Theta(t)$.

Note that in Eq. (\ref{eq:g_behavior}) changing $\sigma_{\chi}\to-\sigma_{\chi}$
is equivalent to interchanging the positive and negative modes, therefore
we can assume $\sigma_{\chi}>0$ in this paper without loss of generality.

\section{Conditions for CKW state}

\label{sec:cond}In this section we will study the evolution of the
fields in the basis of VSH and look for the conditions for the CKW
state.

From Eq. (\ref{eq:cos_def}) we obtain $\cos^{2}\theta(t)$ in VSH,
\begin{equation}
\cos^{2}\theta(t)=\frac{\left\{ \int dk\, k^{3}\left[g_{+}(t,k)-g_{-}(t,k)\right]\right\} ^{2}}{\int dk\, k^{4}\left[g_{+}(t,k)+g_{-}(t,k)\right]\times\int dk\, k^{2}\left[g_{+}(t,k)+g_{-}(t,k)\right]}.\label{eq:cos_g}
\end{equation}
To verify $\cos^{2}\theta(t)<1$, it is better to rewrite $\cos^{2}\theta(t)$
in a more symmetric form, 
\begin{equation}
\cos^{2}\theta(t)=\frac{\int_{0}^{\infty}dk_{1}dk_{2}\,\left(2k_{1}^{3}k_{2}^{3}\right)\times\left[g_{+}(t,k_{1})-g_{-}(t,k_{1})\right]\left[g_{+}(t,k_{2})-g_{-}(t,k_{2})\right]}{\int_{0}^{\infty}dk_{1}dk_{2}\,\left(k_{1}^{4}k_{2}^{2}+k_{1}^{2}k_{2}^{4}\right)\times\left[g_{+}(t,k_{1})+g_{-}(t,k_{1})\right]\left[g_{+}(t,k_{2})+g_{-}(t,k_{2})\right]}.
\end{equation}
The difference between the denominator and numerator is 
\begin{equation}
\mathrm{Diff}\geq8\int_{0}^{\infty}dk_{1}dk_{2}\, k_{1}^{3}k_{2}^{3}g_{+}(t,k_{1})g_{-}(t,k_{2})\geq0,\label{eq:diff}
\end{equation}
where we have used $k_{1}^{4}k_{2}^{2}+k_{1}^{2}k_{2}^{4}\geq2k_{1}^{3}k_{2}^{3}$
in the first inequality. From the inequality (\ref{eq:diff}) it is
obvious that $\cos^{2}\theta(t)\leq1$, where the equality holds for
$k_{1}=k_{2}$ and one of $g_{+}(t,k)$ and $g_{-}(t,k)$ is zero.

Therefore we have two conditions under which $\lim_{t\to\infty}\cos^{2}\theta(t)=1$
is satisfied: 
\begin{align}
(1) & g_{-}(t,k)\to0,\text{ leaving only }g_{+}(t,k),\nonumber \\
(2) & g_{+}(t,k)\to\delta(k-k_{\text{c}}),\label{eq:condition-old}
\end{align}
where $k_{\text{c}}(t)$ is the central momentum of $g_{+}(t,k)$
during evolution. Both conditions can be physically understood. The
first condition is actually the presence of $\sigma_{\chi}$ (we have
assumed $\sigma_{\chi}>0$), which makes positive modes grow with
time while negative modes decay away. It means that the CKW state
should contain only positive (or negative) helicity mode only, which
is reasonable because the CKW state is the eigenstate of the curl
operator. For the second condition, we notice that $\mathbf{W}_{lm}^{\pm}$
bases themselves are CKW states from Eq. (\ref{eq:w_curl}), therefore
one single mode in the expansion (\ref{eq:Bj_expand}) is natually
the CKW state. The authors of Ref. \cite{Hirono:2015rla} observed
$g_{+}(t,k)\to\delta(k-k_{\text{c}})$ in the evolution to the CKW
state. However, the delta function is not well defined mathematically,
so the second condition is hard to implement and we must find a better
one to replace it.

With the solutions for $g_{\pm}(t,k)$ in Eq. (\ref{eq:g_behavior}),
we obtain the time behavior of $\cos^{2}\theta(t)$ from Eq. (\ref{eq:cos_g})
\begin{equation}
\cos^{2}\theta(t)=\frac{\left[I_{3}^{+}(t)-I_{3}^{-}(t)\right]^{2}}{\left[I_{4}^{+}(t)+I_{4}^{-}(t)\right]\left[I_{2}^{+}(t)+I_{2}^{-}(t)\right]},\label{eq:cos_I}
\end{equation}
where the time functions $I_{n}^{\pm}(t)$ are defined by 
\begin{equation}
I_{n}^{\pm}(t)=\int_{0}^{\infty}dk\, k^{n}g_{\pm}(t_{0},k)e^{-2k^{2}\Lambda(t)\pm2k\Theta(t)}.
\end{equation}
Here $n=2,3,4$ are the powers of $k$ in the integrals for $\langle\mathbf{B},\mathbf{B}\rangle$,
$\langle\mathbf{j},\mathbf{B}\rangle$ and $\langle\mathbf{j},\mathbf{j}\rangle$,
respectively.

If the initial spectrum functions $g_{\pm}(t_{0},k)$ contain only
$k^{r}$ ($r$ is a real number), $e^{-2C_{1}k}$ or $e^{-C_{2}^{2}k^{2}}$
($C_{1}$ and $C_{2}$ are real constants), the integrals $I_{n}^{\pm}$
are just Gaussian-like integrals and easy to deal with. In this paper
we assume that $g_{\pm}(t_{0},k)$ take the following form 
\begin{equation}
g_{\pm}(t_{0},k)\sim k^{r}e^{-2C_{1}k-C_{2}^{2}k^{2}}.\label{eq:g_t0}
\end{equation}
A typical example of magnetic fields expressed in such a form is the
Hopf state \cite{irvine2008linked}. Although this assumption narrows
the scope of $g_{\pm}(t_{0},k)$, it is still general enough: these
three kinds of functions are widely used in other fields of physics.
It is natural to combine $g_{\pm}(t_{0},k)$ with $k^{n}e^{-2k^{2}\Lambda(t)\pm2k\Theta(t)}$
in the integrand of $I_{n}^{\pm}(t)$ and re-define the time function
$I_{r+n}^{\pm}(t)$ as 
\begin{equation}
I_{r+n}^{\pm}(t)\equiv\int_{0}^{\infty}dk\, k^{r+n}\exp\left\{ -\left[2\Lambda(t)+C_{2}^{2}\right]k^{2}+2\left[-C_{1}\pm\Theta(t)\right]k\right\} ,
\end{equation}
where $r$ is the power of $k$ in the initial spectrum functions
$g_{\pm}(t_{0},k)$. Note that $I_{-1}^{\pm}$ does not converge,
so we assume $r+n>-1$. By changing the integral variable $k\to\kappa=k\sqrt{2\Lambda(t)+C_{2}^{2}}$
where $2\Lambda(t)+C_{2}^{2}$ is always positive by definition, we
can rewrite $I_{r+n}^{\pm}(t)$ in the form 
\begin{equation}
I_{r+n}^{\pm}(t)=\frac{G_{r+n}(a^{\pm})}{\left[2\Lambda(t)+C_{2}^{2}\right]^{(r+n+1)/2}},
\end{equation}
where $G_{r+n}(a^{\pm})$ are defined by 
\begin{equation}
G_{r+n}(a^{\pm})\equiv\int_{0}^{\infty}d\kappa\,\kappa^{r+n}e^{-\kappa^{2}+2a^{\pm}\kappa},\label{eq:G_def}
\end{equation}
with the time functions $a^{\pm}(t)$ by 
\begin{equation}
a^{\pm}(t)=\frac{\pm\Theta(t)-C_{1}}{\sqrt{2\Lambda(t)+C_{2}^{2}}}.\label{eq:a_def}
\end{equation}
We rewrite $\cos^{2}\theta(t)$ in Eq. (\ref{eq:cos_I}) as a function
of $a^{\pm}(t)$ through $G_{r+n}(a^{\pm})$, 
\begin{equation}
\cos^{2}\theta(t)=\frac{\left[G_{r+3}(a^{+})-G_{r+3}(a^{-})\right]^{2}}{\left[G_{r+4}(a^{+})+G_{r+4}(a^{-})\right]\left[G_{r+2}(a^{+})+G_{r+2}(a^{-})\right]}.\label{eq:cos_Grn}
\end{equation}

We give relevant properties of $G_{r+n}(a)$ in Appendix \ref{sec:property}.
One property is that $G_{r+n}(a)$ are monotonically increasing functions
of $a$, which approach zero at $a\to-\infty$, but rise sharply to
$+\infty$ at $a\to+\infty$. By Eq. (\ref{eq:a_def}), if $\Theta(t)$
grows faster than $\sqrt{\Lambda(t)}$ with $t$, we have $a^{\pm}(t)\rightarrow\pm\infty$
as $t\rightarrow+\infty$. As time goes on, $G_{r+n}(a^{+})$ associated
with positive modes will grow up but $G_{r+n}(a^{-})$ associated
with negative modes will decay away. At $t\to+\infty$, Eq. (\ref{eq:cos_Grn})
becomes 
\begin{equation}
\cos^{2}\theta(t)\approx\frac{\left[G_{r+3}(a^{+})\right]^{2}}{G_{r+4}(a^{+})G_{r+2}(a^{+})}.\label{eq:cos_Grn_p}
\end{equation}
This fulfills the first condition for the CKW state in (\ref{eq:condition-old}),
i.e. only the positive modes survive at the end of the time evolution.

We also show in Appendix \ref{sec:property} that the right hand side
of Eq. (\ref{eq:cos_Grn_p}) tend to 1 if and only if $a^{+}\to+\infty$.
To make $\cos^{2}\theta(t)\to1$ at $t\to+\infty$ requires $a^{+}(t)\to+\infty$,
or 
\begin{equation}
\lim_{t\to\infty}\frac{\Theta(t)-C_{1}}{\sqrt{2\Lambda(t)+C_{2}^{2}}}\to+\infty.
\end{equation}
So we can summarize the conditions for the CKW state to be reached
in time evolution: 
\begin{align}
(1) & \sigma_{\chi}\neq0,\nonumber \\
(2) & \lim_{t\to\infty}\frac{\Theta(t)}{\sqrt{\Lambda(t)}}\to+\infty.\label{eq:condition}
\end{align}
Note that $\sigma_{\chi}$ or $\Theta(t)$ plays an essential role:
it makes negative modes more and more suppressed while making positive
modes blow up as time goes on. At the same time it makes $a^{+}\to+\infty$
at $t\to+\infty$ so that $\cos^{2}\theta(t)\to1$.

In heavy-ion collisions, $\sigma$ and $\sigma_{\chi}$ are decreasing
functions of $t$ as the result of the expansion of the QGP matter.
It is natural to assume that $\sigma$ and $\sigma_{\chi}$ fall with
time in power laws \cite{Tuchin:2013ie}, $\sigma(t)\sim t^{-\alpha}$
and $\sigma_{\chi}(t)\sim t^{-\beta}$, where $\alpha,\beta>0$. This
can be justified by the fact that $\sigma\sim5.8\frac{T}{T_{c}}$
MeV \cite{Ding:2010ga} and $\sigma_{\chi}\sim\mu_{5}$ \cite{Kharzeev:2009pj},
where both the temperature $T$ and the chiral chemical potential
$\mu_{5}$ decrease with time in power laws in expansion. In this
case we have $\Lambda(t)\sim t^{1+\alpha}$, $\Theta(t)\sim t^{1+\alpha-\beta}$
following Eq. (\ref{eq:lambda_theta}), and $a^{+}\sim t^{(1+\alpha)/2-\beta}$,
the condition for the CKW state now becomes 
\begin{equation}
\sigma_{\chi}\neq0,\text{ and }\beta<\frac{1+\alpha}{2}.\label{eq:condition-pow-law}
\end{equation}
The above condition is very easy to check and it is one of the most
useful and practical criteria in this paper. 

The fact that a large $\sigma_{\chi}$ will bring the system to the
CKW state shows that a non-Ohmic current may play a crucial role in
the process of reaching the CKW state. This suggests that in classical
plasmas systems, a non-Ohmic current, e.g., the Hall current, could
produce the same effect. In a classical system with the Hall current
and negligible flow velocity, the evolution of the magnetic field
is governed by 
\begin{equation}
\sigma\frac{\partial\mathbf{B}}{\partial t}=\nabla^{2}\mathbf{B}-\frac{\sigma}{ne}\nabla\times(\mathbf{j\times B}),
\end{equation}
where $n$ is the density of the plasma, $e$ is electron charge,
and the last is the Hall current term. Obviously, with a small resistivity,
the system approaches equilibrium when the CKW state is reached.

\section{Examples and tests of conditions}

\label{sec:examples}In this section we will look at examples of the
CKW state to test the conditions we propose in the last section.

\subsection{With $g_{+}$ only}

As the first example, let us consider an initial spectrum with only
$g_{+}(t_{0},k)$ without $g_{-}(t_{0},k)$. We assume $g_{+}(t_{0},k)$
has the following form, 
\begin{equation}
g_{+}(t_{0},k)=4H_{0}L^{3}k\, e^{-2Lk},\label{eq:g_t0_p}
\end{equation}
where $L$ characterizes the length scale of the magnetic field, $H_{0}$
is the initial magnetic helicity. The normalization constant is chosen
to be $4H_{0}L^{3}$ which gives the initial magnetic helicity of
the spectrum, 
\begin{equation}
H_{0}=\int_{0}^{\infty}dk\, kg_{+}(t_{0},k).\label{eq:H0_p}
\end{equation}
From Eq. (\ref{eq:cos_Grn_p}) we obtain 
\begin{equation}
\cos^{2}\theta(t)=\frac{\left[G_{4}(a)\right]^{2}}{G_{5}(a)G_{3}(a)},\label{eq:cos_G_p}
\end{equation}
where 
\begin{equation}
a(t)=\frac{\Theta(t)-L}{\sqrt{2\Lambda(t)}}\label{eq:a_p}
\end{equation}
is given by Eq. (\ref{eq:a_def}) with $C_{1}=L$ and $C_{2}=0$.
Here we have suppressed the superscript of $a^{+}$ and simply denote
$a\equiv a^{+}$.

To verify our conditions for the CKW state, we consider following
cases: 
\begin{align}
\text{a)} & \;\sigma=\sigma_{0},\ \sigma_{\chi}=\sigma_{\chi0}\nonumber \\
\text{b)} & \;\sigma=\sigma_{0},\ \sigma_{\chi}=0\nonumber \\
\text{c)} & \;\sigma=\sigma_{0},\ \sigma_{\chi}=\sigma_{\chi0}(t_{0}/t)^{1/2}\nonumber \\
\text{d)} & \;\sigma=\sigma_{0}(t_{0}/t)^{1/3},\ \sigma_{\chi}=\sigma_{\chi0}(t_{0}/t)^{1/2}\label{eq:four-cases}
\end{align}
In case a) both $\sigma$ and $\sigma_{\chi}$ are constants, which
is used in Refs. \cite{Tuchin:2014iua,Li:2016tel} to calculate the
magnetic field in medium. In this case, we have $\alpha=\beta=0$
and $a(t)\sim t^{1/2}$ in late time which satisfies the condition
$\beta<\frac{1+\alpha}{2}$, and we can see the effect of non-vanishing
constant $\sigma_{\chi}$. Such an effect can be seen by comparing
with case b) in which we switch off $\sigma_{\chi}$. In case c) $\sigma$
is still a constant as same as case a), but $\sigma_{\chi}\sim t^{-1/2}$
is chosen to break the condition $\beta<\frac{1+\alpha}{2}$ with
$\alpha=0$ and $\beta=1/2$. In case d), the values $\alpha=1/3$
and $\beta=1/2$ are used in Refs. \cite{Tuchin:2013ie,Yamamoto:2016xtu},
which are thought to be more reasonable in heavy-ion collisions. But
we note that in real situations of heavy-ion collisions, the time
behaviors of $\sigma$ and $\sigma_{\chi}$ can be very complicated
(may not follow power laws), but our conditions in (\ref{eq:condition})
are still applicable. 

\begin{figure}[h]
\begin{centering}
\includegraphics[width=0.5\linewidth]{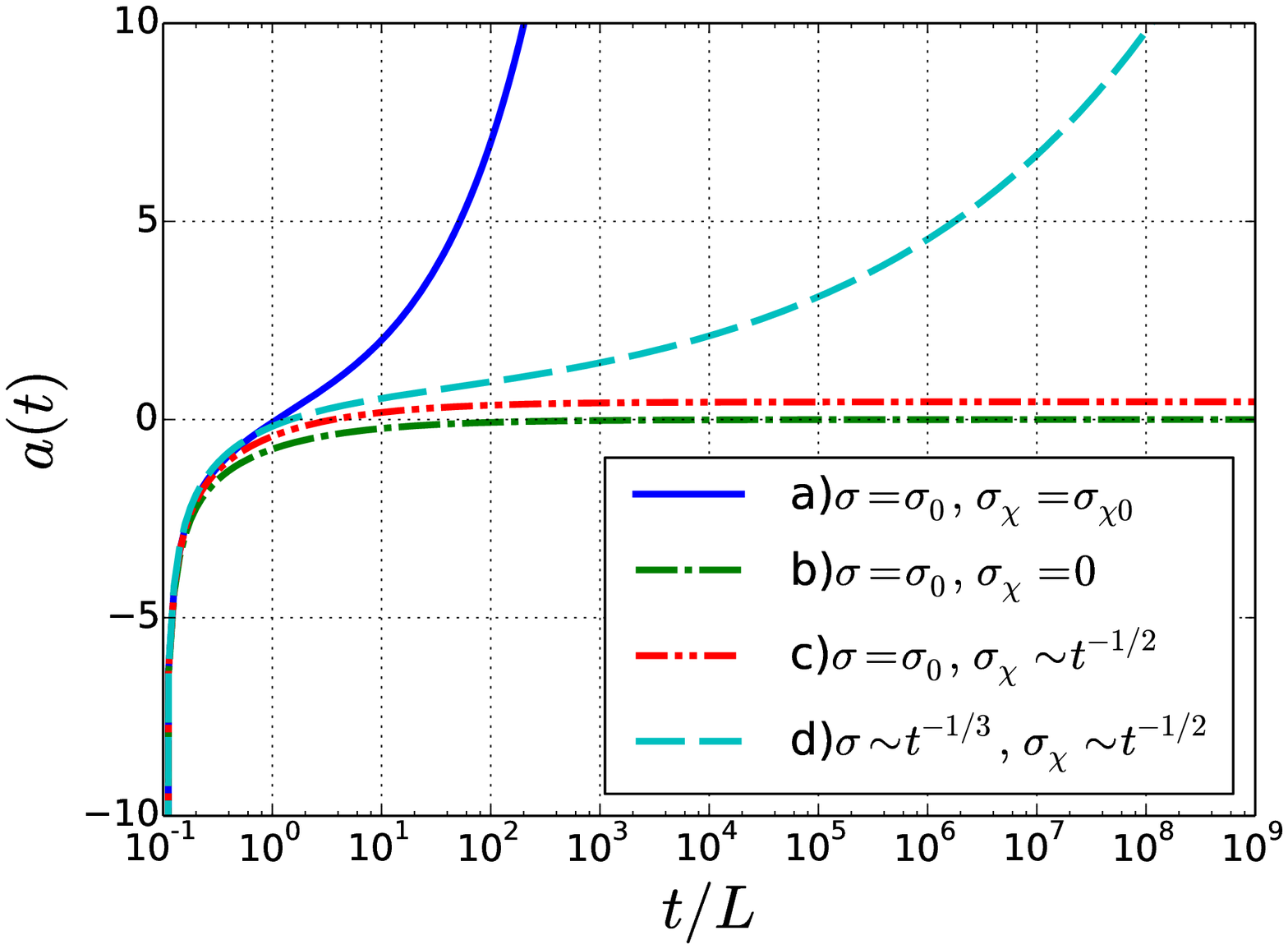}\includegraphics[width=0.5\linewidth]{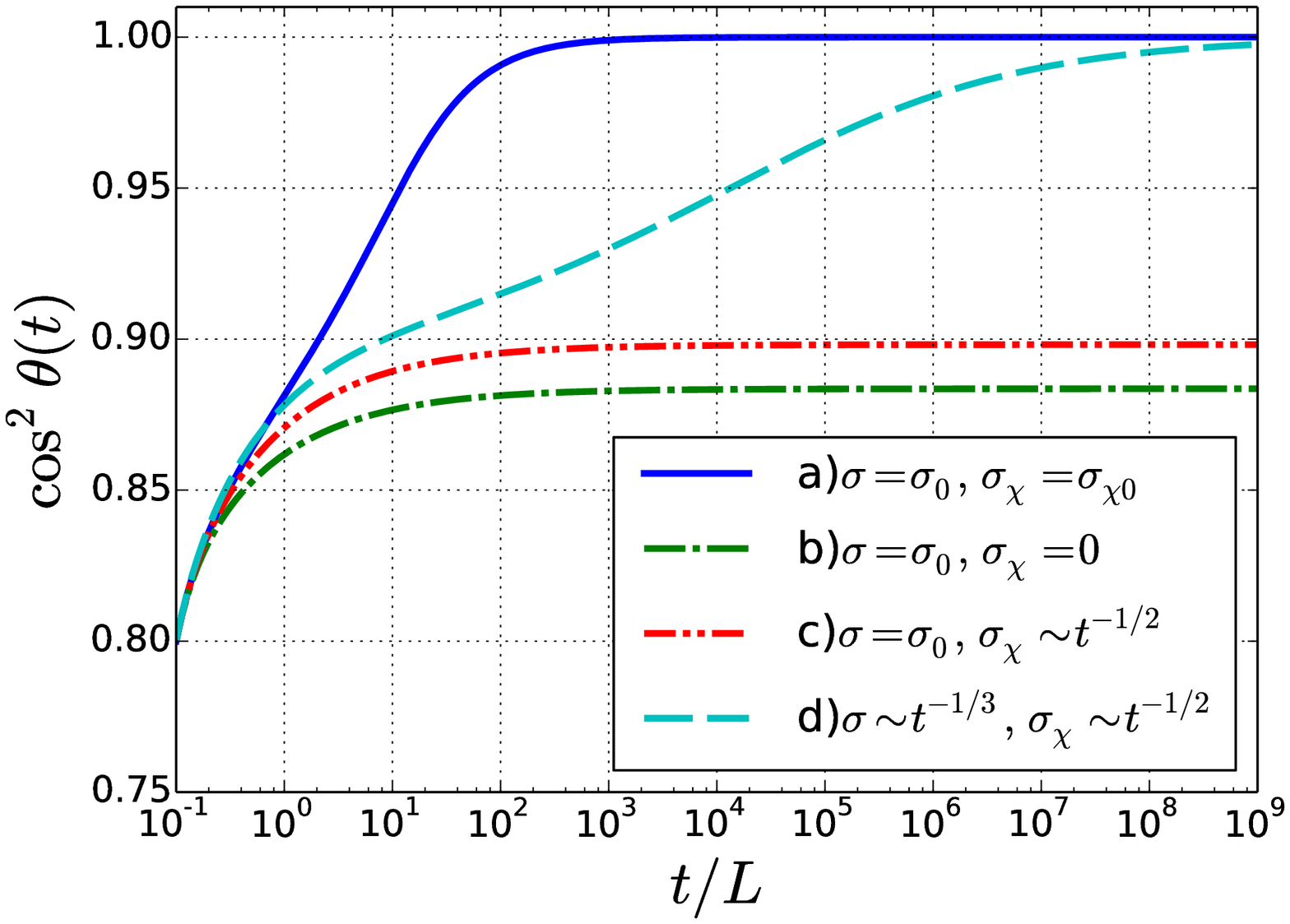} 
\par\end{centering}

\caption{\label{fig:cos1}(color online) The time functions of $a(t)$ (left
panel) and $\cos^{2}\theta(t)$ (right panel) in different cases.
In case a) (blue solid) and d) (cyan dashed), $a(t)$ increase toward
$+\infty$ and $\cos^{2}\theta(t)$ tend toward 1. In case b) (green
dot-dashed) and c) (red dot-dot-dashed), $a(t)$ tend toward 0 and
$\frac{\sigma_{\chi0}}{\sigma_{0}}\sqrt{2t_{0}\sigma_{0}}$ respectively,
and $\cos^{2}\theta(t)$ tend toward saturation values less than 1. }
\end{figure}

For numerical simulation, we choose $\sigma_{0}=\sigma_{\chi0}=1/L$
and $t_{0}=0.1L$. The results are shown in Fig. \ref{fig:cos1}.
Indeed in case a) and d), the CKW state can be reached. As $t$ goes
from $t_{0}$ to $\infty$, according to Eqs. (\ref{eq:cos_G_p},
\ref{eq:a_p}, \ref{eq:GGG_limit}), $a(t)$ evolves from $-\infty$
to $\infty$, and $\cos^{2}\theta(t)$ evolves from 0.8 to 1. In case
b) and c) the condition (\ref{eq:condition-pow-law}) is not satisfied,
the CKW state is inaccessible. Indeed the simulation shows that it
is true since $a(t)$ tends toward 0 and $\frac{\sigma_{\chi0}}{\sigma_{0}}\sqrt{2t_{0}\sigma_{0}}$
in case b) and c) at $t\to\infty$, respectively. Even though $a(t)$
and $\cos^{2}\theta(t)$ increase with $t$, we have $\cos^{2}\theta(t)\to0.884$
and $0.898$ at $t\to+\infty$ corresponding to $a\to0$ and $a\to\frac{\sigma_{\chi0}}{\sigma_{0}}\sqrt{2t_{0}\sigma_{0}}$
respectively. All these results show that the conditions work well.

\begin{figure}[h]
\begin{centering}
\includegraphics[width=0.5\linewidth]{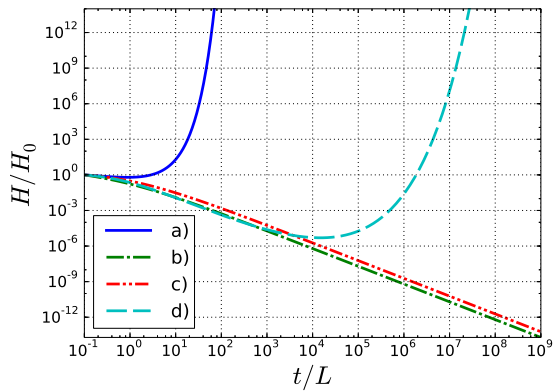} 
\par\end{centering}

\caption{\label{fig:helicity}(color online) The time dependence of the magnetic
helicity $H$ in different cases. In case a) (blue solid) and d) (cyan
dashed), $H$ blow up at some time. In case b) (green dot-dashed)
and c) (red dot-dot-dashed), $H$ decay to zero in late time. }
\end{figure}

But we should point out that constant $\sigma$ and $\sigma_{\chi}$
or even the power law decayed $\sigma$ and $\sigma_{\chi}$ may not
be physical since once $\sigma_{\chi}$ persists for a long time,
$\Theta(t)$ growing faster than $\sqrt{\Lambda(t)}$ will make some
physical quantities diverge. We look at the magnetic helicity $H$
and the magnetic energy $W$, 
\begin{gather}
H=\int k=4H_{0}L^{3}\frac{G_{2}(a)}{\left[2\Lambda(t)\right]^{3/2}},\nonumber \\
W=\frac{1}{2}\int k^{2}=2H_{0}L^{3}\frac{G_{3}(a)}{\left[2\Lambda(t)\right]^{2}}.\label{eq:HW}
\end{gather}
The numerical results of the magnetic helicity are shown in Fig. \ref{fig:helicity}.
The results of the magnetic energy are similar. In case b) and c),
since $a(t)$ and $G_{n}(a)$ converge to constants, but $\Lambda(t)$
keeps growing, both $H$ and $W$ finally decay to zero following
Eq. (\ref{eq:HW}). However in case a) and d), $a(t)$ and $G_{n}(a)$
increase to $+\infty$ in late time, and from Eq (\ref{eq:G_limit})
$G_{n}(a)\sim a^{n}\exp(a^{2})$ grow much faster than $\left[\Lambda(t)\right]^{(n+1)/2}$
to make $H$ and $W$ blow up.

From Eq. (\ref{eq:d_g/dt}), we see that the spectrum $g_{+}(t,k)$
grows exponentially in time for $k<\sigma_{\chi}$, such an instability
has been discussed in \cite{Tuchin:2014iua,Manuel:2015zpa}, see also
\cite{Akamatsu:2013pjd}. This instability is the source of the divergence
of $H$ and $W$. Such an unphysical inflation can be understood:
the appearance of $\sigma_{\chi}\mathbf{B}$ in the induced current
leads to the positive feedback that the magnetic field itself induces
the magnetic field. If we put no constraint on $\sigma_{\chi}$, as
the result, the magnetic field will keep growing and finally blow
up at some time. This of course breaks conservation laws. One way
to avoid such divergences is to implement conservation laws in the
system. This is the topic of the next subsection.

\subsection{With $g_{+}$ only and dynamical $\sigma_{\chi}$}

We now consider imposing the total helicity conservation in Eq. (\ref{eq:conservation}).
This has been implemented in Ref. \cite{Manuel:2015zpa,Hirono:2015rla}.
Here we focus on the approach to the CKW state in evolution. For simplicity,
we can parameterize $\sigma_{\chi}$ as 
\begin{equation}
\sigma_{\chi}(t)=k_{h}\left[H_{\text{total}}-H(t)\right],\label{eq:seesaw}
\end{equation}
where $k_{h}$ and $H_{\text{total}}$ (total helicity) are constants.
From Eq. (\ref{eq:seesaw}), we see that the requirement $\sigma_{\chi}>0$
leads to $H<H_{\text{total}}$. The initial spectrum $g_{+}(t_{0},k)$
is assumed to be the same as Eq. (\ref{eq:g_t0_p}), so we have $a(t)=\frac{\Theta(t)-L}{\sqrt{2\Lambda(t)}}$.
The parameters are chosen to be $\sigma=k_{h}=1/L$, $t_{0}=0.1L$,
and $H_{0}/H_{\text{total}}=3/5$, where $H_{0}$ is given by Eq.
(\ref{eq:H0_p}). Since $\sigma$ is a constant, we have $\Lambda(t)=\frac{1}{\sigma}(t-t_{0})$.
We can solve $\sigma_{\chi}(t)$ self-consistently through $\Theta(t)$,
\begin{equation}
\frac{d}{dt}\Theta(t)=\frac{k_{h}}{\sigma}\left\{ H_{\text{total}}-H[\Theta(t)]\right\} ,\label{eq:h-theta}
\end{equation}
where we have used $\sigma_{\chi}(t)=\sigma\frac{d}{dt}\Theta(t)$
and that $H$ depends on $\Theta(t)$ through $a(t)$ in Eq. (\ref{eq:HW}).

\begin{figure}[h]
\begin{centering}
\includegraphics[width=0.5\linewidth]{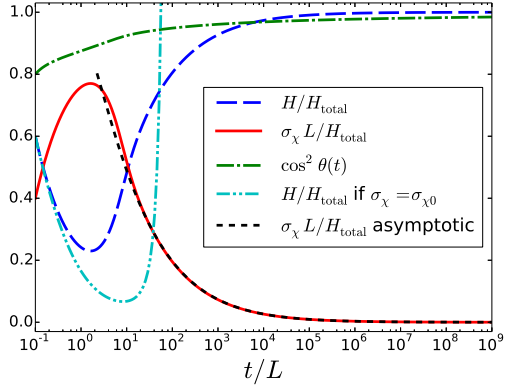} 
\par\end{centering}

\caption{\label{fig:cosH}(color online) The magnetic helicity $H$ (blue dashed),
chiral magnetic conductivity $\sigma_{\chi}$ (red solid) and $\cos^{2}\theta(t)$
(green dot-dashed) under the constraint $\sigma_{\chi}L+H=H_{\text{total}}$
as functions of $t$. The curve of $H$ (cyan dot-dot-dashed) is the
result of constant chiral magnetic conductivity $\sigma_{\chi}=\sigma_{\chi0}$.
The black short dashed curve is the result by Eq. (\ref{eq:sigma_chi_large t}).
As $\cos^{2}\theta(t)$ approaches the limit 1, $H$ takes all share
of $H_{\text{total}}$, and $\sigma_{\chi}$ tends to decrease as
Eq. (\ref{eq:sigma_chi_large t}) asymptotically. }
\end{figure}

The numerical results for $\sigma_{\chi}(t)$, $H(t)$ and $\cos^{2}\theta(t)$
are presented in Fig. \ref{fig:cosH}. For comparison, we also show
the result for constant $\sigma_{\chi}$ with $\sigma_{\chi}=\sigma_{\chi0}=2H_{\text{total}}/(5L)$.
In both cases, at the beginning, $a(t)=\frac{\Theta(t)-L}{\sqrt{2\Lambda(t)}}$
is not large enough to make $G_{2}(a)$ grows faster than $\Lambda(t)^{3/2}$,
which makes $H\sim G_{2}/\Lambda^{3/2}$ in Eq. (\ref{eq:HW}) decrease
with time. After $a$ grows large enough as time goes on, $H$ starts
to increase after reaching a minimum. In the case of dynamical $\sigma_{\chi}$,
according to Eq. (\ref{eq:seesaw}), $H$ and $\sigma_{\chi}$ are
complementary to each other to make up a seesaw system. In this system,
the decreasing of $H$ at the beginning raises the value of $\sigma_{\chi}$
and makes $\Theta(t)$ and $a(t)$ grow faster. As the result, the
turning point comes earlier than the case of constant $\sigma_{\chi}$.
As $H$ keeps growing, $\sigma_{\chi}$ drops down leading to slower
increase of $\Theta(t)$, which makes $H$ grows slower. At the end,
$H$ is saturated to $H_{\text{total}}$ instead of blowing up.

Let us look at the asymptotic time behavior of $\sigma_{\chi}$ as
$a\to\infty$. As the magnetic helicity $H$ is saturated to $H_{\text{total}}$
following Eq. (\ref{eq:HW}), with $G_{2}(a)\sim\sqrt{\pi}a^{2}\exp(a^{2})$
and $\Lambda(t)\sim\frac{1}{\sigma_{0}}t$ at late time, we obtain
\begin{equation}
a^{2}\approx\mathrm{Plog}(ct^{3/2}),
\end{equation}
where $c=\frac{H_{\text{total}}}{\sqrt{2\pi}H_{0}L^{3}\sigma_{0}^{3/2}}$
and $\mathrm{Plog}$ is called product logarithm, which is the inverse
function of $f(x)=xe^{x}$. From $a=\frac{\Theta(t)-L}{\sqrt{2\Lambda(t)}}$,
we obtain at very large $t$, 
\begin{equation}
\Theta(t)\approx\sqrt{\frac{2}{\sigma_{0}}t\mathrm{Plog}(ct^{3/2})}.\label{eq:theta_large t}
\end{equation}
Since the $\mathrm{Plog}$ term increases with $t$, $\Theta(t)$
is always growing faster than $\sqrt{\Lambda(t)}\sim\sqrt{t}$. Thus
the conditions are satisfied and the CKW state can be reached.

Taking a derivative of $\Theta(t)$ with respect to $t$, we obtain
$\sigma_{\chi}(t)$ at late time from Eq. (\ref{eq:theta_large t}),
\begin{equation}
\sigma_{\chi}(t)\approx\sqrt{\frac{\sigma_{0}}{8t}\mathrm{Plog}(ct^{3/2})}\frac{5+2\mathrm{Plog}(ct^{3/2})}{1+\mathrm{Plog}(ct^{3/2})}.\label{eq:sigma_chi_large t}
\end{equation}
The asymptotic behavior of $\sigma_{\chi}(t)$ is $\sigma_{\chi}(t)\approx\sqrt{\frac{\sigma_{0}}{2t}\mathrm{Plog}(ct^{3/2})}$
which decays slower than $t^{-1/2}$. We show the result from Eq.
(\ref{eq:sigma_chi_large t}) in the black short dashed line in Fig.
\ref{fig:cosH}, which agrees with the numerical result very well.

We have also looked at a general spectrum for $g_{+}$ at initial
time, 
\begin{equation}
g_{+}(t_{0},k)=N_{0}k^{r}e^{-2kC_{1}-k^{2}C_{2}^{2}},
\end{equation}
where the normalization constant $N_{0}$ is determined by the initial
magnetic helicity $H_{0}$. We assume $\sigma(t)\sim t^{-\alpha}$
obeying the power law decay in time, which gives $\Lambda(t)\sim t^{1+\alpha}$.
In this case, solving Eq. (\ref{eq:h-theta}) gives the late time
asymptotic behavior, 
\begin{equation}
\Theta(t)\approx\sqrt{(r+1)\Lambda\mathrm{Plog}\left[\frac{4}{r+1}\left(\frac{\sqrt{2}H_{\text{total}}}{\sqrt{\pi}N_{0}}\right)^{2/(r+1)}\Lambda^{(r+2)/(r+1)}\right]}.
\end{equation}
Again we see that $\Theta(t)$ grows faster than $\sqrt{\Lambda(t)}$
and the CKW state can be finally reached.

\subsection{With mixed helicity}

In this example we consider both positive and negative modes. We will
show that only the positive mode survives while the negative mode
decays away in late time. Let us consider the most extreme case in
which the initial spectra of the positive and negative modes are the
same. We take the following initial spectra for $g_{\pm}$, 
\begin{equation}
g_{\pm}(t_{0},k)=4H_{0}L^{3}k\, e^{-2kL}.
\end{equation}
It is obvious that the initial magnetic helicity is zero. Since $g_{\pm}(t_{0},k)\sim k$,
$\cos^{2}\theta(t)$ is given by Eq. (\ref{eq:cos_Grn}) with $r=1$
and $a^{\pm}(t)=\frac{\pm\Theta(t)-L}{\sqrt{2\Lambda(t)}}$. Obviously
at the initial time we have $\cos^{2}\theta(t_{0})=0$ because $a^{+}(t_{0})=a^{-}(t_{0})$.

\begin{figure}[h]
\begin{centering}
\includegraphics[width=0.5\linewidth]{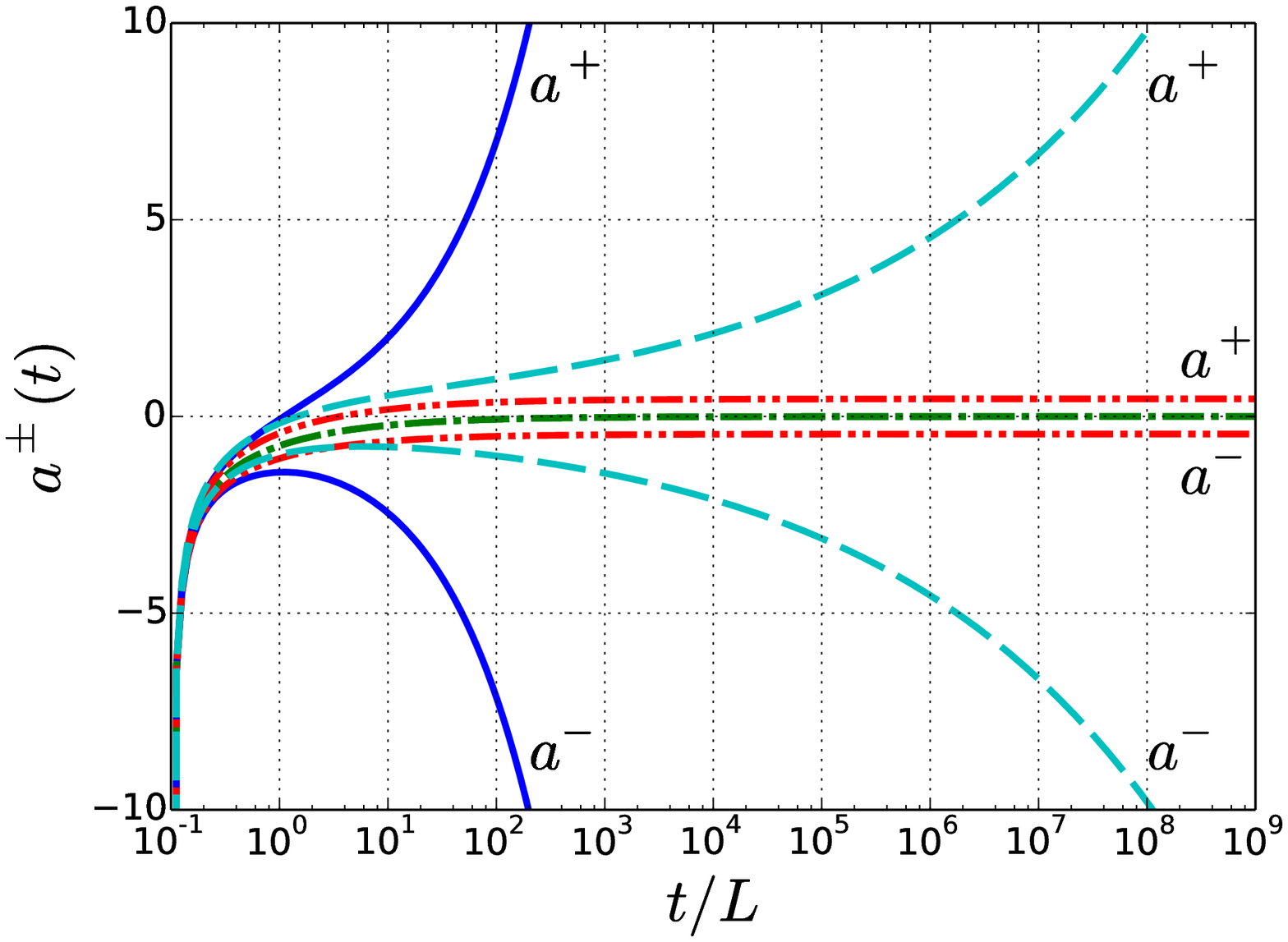}\includegraphics[width=0.5\linewidth]{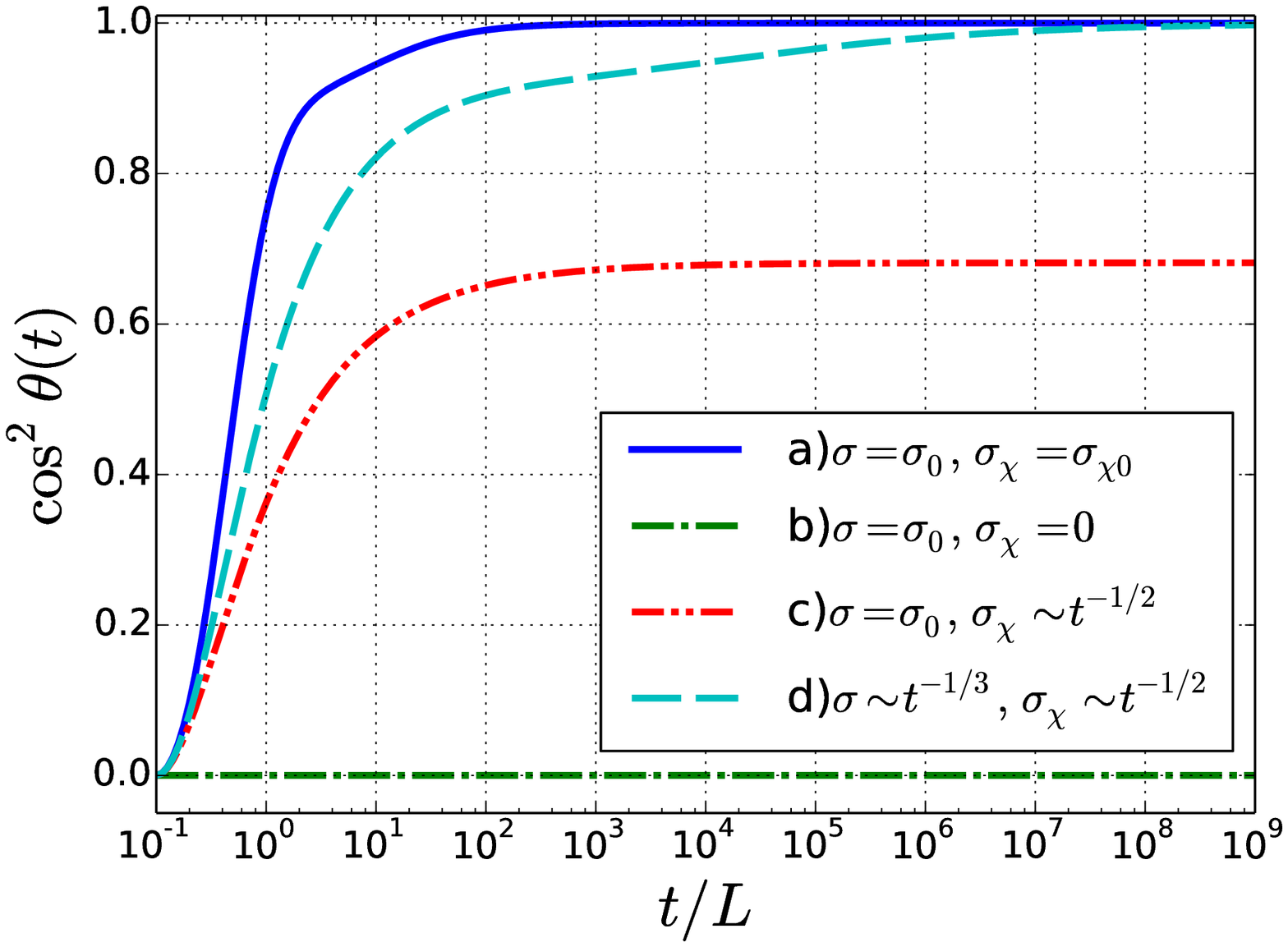} 
\par\end{centering}

\caption{\label{fig:cos2}(color online) The time behaviors of $a^{\pm}(t)$
(left panel) and $\cos^{2}\theta(t)$ (right panel) in different cases.
In case a) (blue solid) and d) (cyan dashed), $\cos^{2}\theta(t)$
tend to 1 in late time. In case c) (red dot-dot-dashed) $\cos^{2}\theta(t)$
tends to a limit less than 1, while in case b) (green dot-dashed)
$\cos^{2}\theta(t)$ is equal to $0$ due to $a^{+}=a^{-}$.}
\end{figure}

We consider again the cases a)-d) as in (\ref{eq:four-cases}). The
results are shown in Fig. \ref{fig:cos2}. The CKW state can be approached
in case a) and d). As is shown in the left panel of Fig. \ref{fig:cos2},
$a^{\pm}(t)$ grow to $\pm\infty$ respectively, therefore at very
late time the negative mode becomes highly suppressed and the positive
mode survives, leading $\cos^{2}\theta(t)$ to approach 1. However
in case b) we have $a^{+}=a^{-}$ and $a^{\pm}\to0$ in late time.
In this case the negative/positive modes are identical in the time
evolution, which results in $\cos^{2}\theta(t)=0$. In case c) we
have $\beta=\frac{1+\alpha}{2}$ which breaks the conditions for the
CKW state. In this case, $a^{\pm}(t)$ do not go to $\pm\infty$ at
$t\to+\infty$, instead they approach the limits $\pm\frac{\sigma_{\chi0}}{\sigma_{0}}\sqrt{2t_{0}\sigma_{0}}$.
We see that $\cos^{2}\theta(t)$ keeps growing with time till it gets
saturation, $\lim_{t\to\infty}\cos^{2}\theta(t)\to0.681$. All these
results are consistent with our expectations.

\section{More general momentum spectra}

\label{sec:polynomial}The power pre-factor $k^{r}$ of initial momentum
spectra $g_{\pm}(t_{0},k)$ in (\ref{eq:g_t0}) can be generalized
to a polynomial, 
\begin{equation}
g_{\pm}(t_{0},k)=\left(\sum_{r=r_{\text{min}}}^{r_{\text{max}}}c_{r}k^{r}\right)e^{-2kC_{1}}e^{-k^{2}C_{2}^{2}},\label{eq:g_t0_poly}
\end{equation}
where $\sum_{r}c_{r}k^{r}>0$ for any $k$ due to the fact that $g_{\pm}$
must be positive. Here $r$ are real numbers with $r>-3$. Ignoring
negative modes, we obtain $\cos^{2}\theta(t)$ from Eq. (\ref{eq:cos_Grn_p}),
\begin{equation}
\cos^{2}\theta(t)=\frac{\left[\sum_{r=r_{\text{min}}}^{r_{\text{max}}}c_{r}G_{r+3}(a)\right]^{2}}{\left[\sum_{r=r_{\text{min}}}^{r_{\text{max}}}c_{r}G_{r+4}(a)\right]\times\left[\sum_{r=r_{\text{min}}}^{r_{\text{max}}}c_{r}G_{r+2}(a)\right]},\label{eq:cos_poly}
\end{equation}
where $a\equiv a^{+}$. According to Eq. (\ref{eq:G_limit}), the
$r_{\text{min}}$-th term is dominant at $a\to-\infty$, while the
$r_{\text{max}}$-th term is dominant at $a\to+\infty$. We can prove
that $\cos^{2}\theta(t)$ in Eq. (\ref{eq:cos_poly}) tends toward
$(r_{\text{min}}+3)/(r_{\text{min}}+4)$ at $a\to-\infty$ by Eq.
(\ref{eq:GGG_limit}). From Eq. (\ref{eq:asymptote}), at large positive
$a$ the ratio $(G_{n})^{2}/(G_{n+1}G_{n-1})$ can be approximated
by $1-1/(2a^{2})$. Hence the right-hand-side of Eq. (\ref{eq:cos_poly})
dominated by $(G_{r_{\text{max}}+3})^{2}/(G_{r_{\text{max}}+4}G_{r_{\text{max}}+2})$
is also approximately $1-1/(2a^{2})$ at large positive $a$ and reaches
1 for $a\to+\infty$. So our conditions for the CKW state are applicable
to the initial momentum spectrum (\ref{eq:g_t0_poly}) with a polynomial
pre-factor.

As examples we consider following initial momentum spectra $g_{+}(t_{0},k)\sim k+k^{2}$,
$k+k^{10}$ and $k-2k^{2}+k^{3}$. We compare them with the power
pre-factor $g_{+}(t_{0},k)\sim k$ we considered in the last section.
We use Eq. (\ref{eq:cos_poly}) for $\cos^{2}\theta(t)$. In all these
cases we have $r_{\text{min}}=1$ and $(r_{\text{min}}+3)/(r_{\text{min}}+4)=0.8$.

\begin{figure}[h]
\begin{centering}
\includegraphics[width=0.5\linewidth]{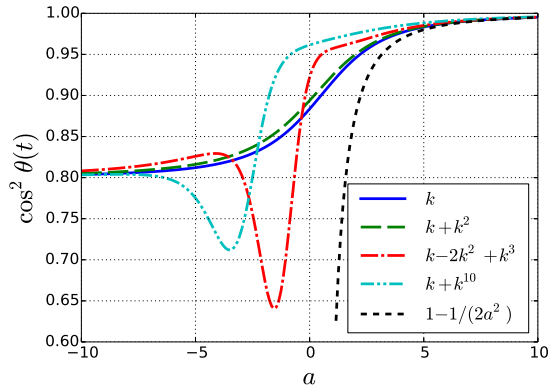} 
\par\end{centering}

\caption{\label{fig:gmix}(color online) The quantity $\cos^{2}\theta(t)$
in Eq. (\ref{eq:cos_poly}) as functions of $a$ for $g_{+}(t_{0},k)\sim k$,
$k+k^{2}$, $k-2k^{2}+k^{3}$ and $k+k^{10}$. They are all near 0.8
at $a\to-\infty$ and fit $1-1/(2a)^{2}$ at large positive $a$.}
\end{figure}

We plot $\cos^{2}\theta(t)$ as functions of $a$ in Fig. \ref{fig:gmix}.
The results show that at $a\to-\infty$, all curves are near 0.8 which
implies the $r_{\text{min}}$-th term is dominant. For moderate values
of $a$, the curves for different initial momentum spectra diverge
significantly. At large positive $a$, they converge again and approach
$1-1/(2a)^{2}$ as expected.

\section{Summary and conclusions}

We investigate the Chandrasekhar-Kendall-Woltjer (CKW) state in a
plasma with currents induced by chiral anomaly in magnetic fields.
We solve the Maxwell-Chern-Simons equations and propose the conditions
at which the CKW state can be finally realized in time evolution.
We decompose the vector potential and the magnetic field into vector
spherical harmonic modes. These modes are labeled by scalar momentum
$k=|\mathbf{k}|$, quantum numbers of orbital angluar momentum $l$
and angular momentum along specific direction $m$, and the photon
(electromagnetic field) helicity $s=\pm$. For each mode, there is
an ordinary differential equation in time whose solution can be easily
found.

We define a quantity $\cos^{2}\theta(t)$ for the CKW state, where
$\theta(t)$ can be regarded as the average angle between $\mathbf{j}$
and $\mathbf{B}$. If $\cos^{2}\theta(t)=1$, the CKW state is reached.
In the real world the CKW state cannot be exactly reached but only
be approached asymptotically, i.e. $\lim_{t\to+\infty}\cos^{2}\theta(t)=1$.
In the bases of vector spherical harmonic functions, a general inner
product of two vector fields can be put into a uniform integral, from
which we can express $\cos^{2}\theta(t)$ in a simple form.

We propose the conditions that the CKW state can be reached in time
evolution for a general class of initial momentum spectra: (i) the
presence of $\sigma_{\chi}$ and (ii) $\Theta(t)$ grows faster in
time than $\sqrt{\Lambda(t)}$, where $\Lambda(t)$ and $\Theta(t)$
are integrations over $t$ of $1/\sigma$ and of $\sigma_{\chi}/\sigma$
respectively.

We take some examples to test these conditions. The numerical results
agree with these conditions very well. In numerical calculations,
we set different values for the parameters $\sigma_{0}$, $\sigma_{\chi0}$,
$t_{0}$, $k_{h}$, $H_{0}/H_{\text{total}}$ and the powers of $k$
in initial momentum spectra. We find that in cases of constant $\sigma_{\chi}$
and some power law decaying functions, although the CKW state can
be approached but the magnetic helicity and magnetic energy will blow
up. Such a problem can be avoided with dynamical $\sigma_{\chi}$
determined self-consistently with a helicity bound and a negative
feedback. The critical decaying behavior of $\sigma_{\chi}$ given
in Eq. (\ref{eq:sigma_chi_large t}) is found both numerically and
analytically. The CKW state can be reached at the end of time evolution.

\textit{Acknowledgments.} QW and XLX thank H. Li and Y.G. Yang for
helpful discussions. QW is supported in part by the Major State Basic
Research Development Program (MSBRD) in China under Grant 2015CB856902
and 2014CB845402 respectively and by the National Natural Science
Foundation of China (NSFC) under the Grant 11535012.

\appendix

\section{Properties of function $G_{n}(a)$}

\label{sec:property}We give some useful properties of the function
$G_{r+n}(a^{\pm})$ in Eq. (\ref{eq:G_def}). For convenience we simply
denote this function as $G_{n}(a)$ in this appendix. We formally
define $G_{n}(a)$ as 
\begin{equation}
G_{n}(a)=\int_{0}^{\infty}d\kappa\,\kappa^{n}e^{-\kappa^{2}+2a\kappa},\label{eq:Gn_def}
\end{equation}
where $n$ is any real number larger than $-1$.

By definition $G_{n}(a)$ can be integrated as 
\begin{equation}
G_{n}(a)=a\Gamma\left(\frac{n+3}{2}\right){}_{1}F_{1}\left(\frac{n+3}{2};\frac{3}{2};a^{2}\right)+\frac{1}{2}\Gamma\left(\frac{n+1}{2}\right){}_{1}F_{1}\left(\frac{n+1}{2};\frac{1}{2};a^{2}\right),\label{eq:G_n}
\end{equation}
where $\Gamma(x)$ is the gamma function and $_{1}F_{1}(a;b;x)$ is
the confluent hypergeometric function of the first kind. The above
formula is applicable for any real number $n>-1$. For $n$ being
integer, $G_{n}(a)$ can also be determined recursively by 
\begin{equation}
G_{n}(a)=aG_{n-1}(a)+\frac{n-1}{2}G_{n-2}(a),
\end{equation}
with 
\begin{align}
G_{0}(a) & =\frac{\sqrt{\pi}}{2}\exp(a^{2})\mathrm{erfc}(-a),\nonumber \\
G_{1}(a) & =\frac{1}{2}\left[1+\sqrt{\pi}a\exp(a^{2})\mathrm{erfc}(-a)\right],
\end{align}
where $\mathrm{erfc}(x)$ is the complementary error function, $\mathrm{erfc}(x)=\frac{2}{\sqrt{\pi}}\int_{x}^{\infty}dt\exp(-t^{2})$.

\begin{figure}[h]
\begin{centering}
\includegraphics[width=0.5\linewidth]{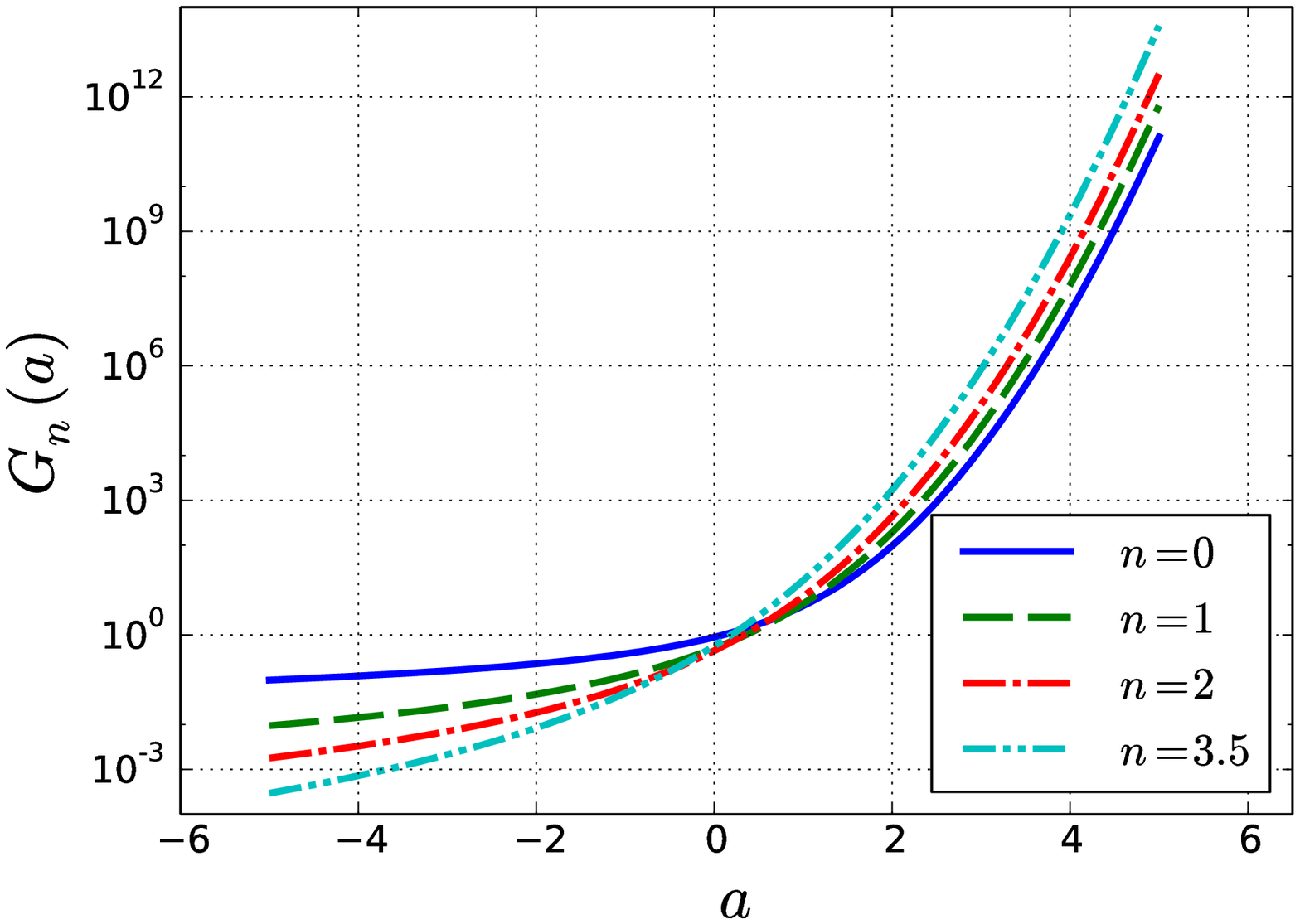}\includegraphics[width=0.5\linewidth]{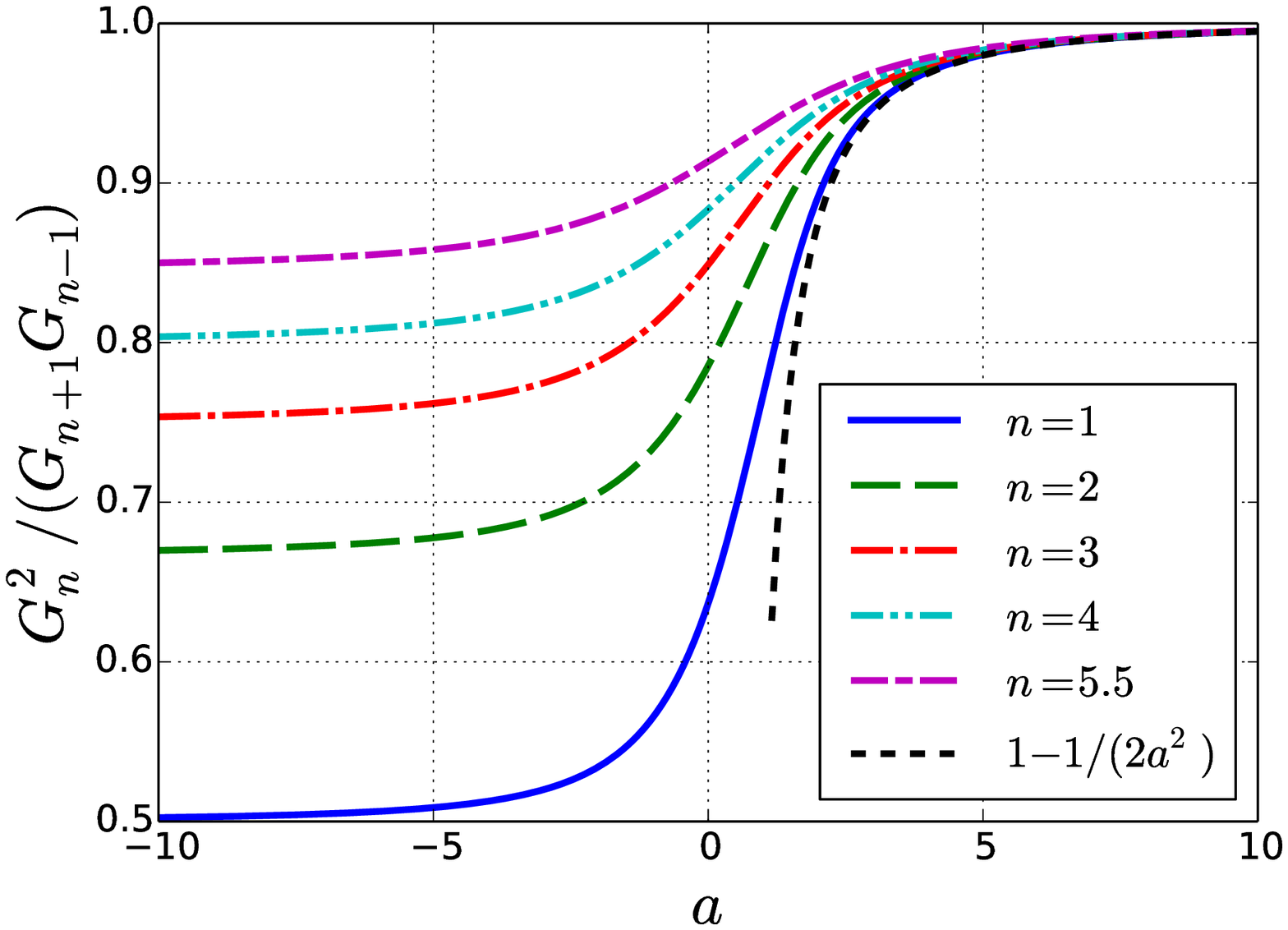} 
\par\end{centering}

\caption{\label{fig:gggg}(color online) Left panel: $G_{n}(a)$ as functions
of $a$ for $n=0,1,2,3.5$. They tend to 0 at $a\to-\infty$ and $+\infty$
at $a\to+\infty$. Right panel: $[G_{n}(a)]^{2}/[G_{n-1}(a)G_{n+1}(a)]$
as functions of $a$. They increase from $\frac{n}{n+1}$ to 1 in
the range $a\in(-\infty,+\infty)$ and fit asymptote $1-1/(2a)^{2}$
for large positive $a$.}
\end{figure}

From the definition Eq. (\ref{eq:Gn_def}), it is straightforward
to have 
\begin{equation}
\frac{d}{da}G_{n}(a)=2G_{n+1}(a).
\end{equation}
Since $G_{n+1}(a)$ are always positive, $G_{n}(a)$ are monotonically
increasing functions of $a$. We plot $G_{n}(a)$ in Fig. \ref{fig:gggg}
for $n$ being integers $0,1,2$ and a real number $3.5$ as examples.
They all approach zero at $a\to-\infty$, but rise sharply to $+\infty$
at $a\to+\infty$.

For the purpose of this paper, we are concerned with the asymptotic
behavior of $G_{n}(a)$ at $a\to\infty$. From Eq. (\ref{eq:G_n})
we have found that 
\begin{equation}
\lim_{a\to+\infty}\frac{G_{n}(a)}{\sqrt{\pi}a^{n}\exp(a^{2})}=\lim_{a\to-\infty}\frac{(-2a)^{n+1}}{\Gamma(n+1)}G_{n}(a)=1.\label{eq:G_limit}
\end{equation}
This means $G_{n}(a)$ approach $\sqrt{\pi}a^{n}\exp(a^{2})$ and
$\Gamma(n+1)\left(-\frac{1}{2a}\right)^{n+1}$ at $a\to\pm\infty$
respectively, and implies that 
\begin{equation}
\lim_{a\to+\infty}\frac{\left[G_{n}(a)\right]^{2}}{G_{n+1}(a)G_{n-1}(a)}=1,\quad\lim_{a\to-\infty}\frac{\left[G_{n}(a)\right]^{2}}{G_{n+1}(a)G_{n-1}(a)}=\frac{n}{n+1}.\label{eq:GGG_limit}
\end{equation}
We plot the quantity $\left[G_{n}(a)\right]^{2}/\left[G_{n+1}(a)G_{n-1}(a)\right]$
in Fig. \ref{fig:gggg} for $n=1,2,3,4,5.5$. It shows that they increase
with $a$ in S-shape from $\frac{n}{n+1}$ to 1 in the range $a\in(-\infty,+\infty)$.

To obtain the asymptote of $\left[G_{n}(a)\right]^{2}/\left[G_{n+1}(a)G_{n-1}(a)\right]$
at $a\to+\infty$, we expand $G_{n}(a)$ for large $a$ to the next
to leading order in $1/a$, 
\begin{equation}
G_{n}(a)\approx\sqrt{\pi}a^{n}\left[1+\frac{n(n-1)}{4a^{2}}\right]\exp(a^{2}).
\end{equation}
So the asymptote of the the ratio $\left[G_{n}(a)\right]^{2}/\left[G_{n+1}(a)G_{n-1}(a)\right]$
is 
\begin{align}
\frac{\left[G_{n}(a)\right]^{2}}{G_{n+1}(a)G_{n-1}(a)} & \approx\frac{\left[1+n(n-1)/(4a^{2})\right]^{2}}{\left[1+n(n+1)/(4a^{2})\right]\left[1+(n-1)(n-2)/(4a^{2})\right]}\nonumber \\
 & \approx\frac{1+(n^{2}-n)/(2a^{2})}{1+(n^{2}-n+1)/(2a^{2})}\nonumber \\
 & =1-\frac{1}{2a^{2}+n^{2}-n+1}\nonumber \\
 & \approx1-\frac{1}{2a^{2}}.\label{eq:asymptote}
\end{align}
Fig. \ref{fig:gggg} shows that all the quantity $\left[G_{n}(a)\right]^{2}/\left[G_{n+1}(a)G_{n-1}(a)\right]$
approach the asymptote $1-1/(2a)^{2}$ for large $a$. We see with
the larger $n^{2}-n+1$, they deviate the asymptote further. At very
large $a$, they fit $1-1/(2a)^{2}$ very well. We also see $\left[G_{n}(a)\right]^{2}/\left[G_{n+1}(a)G_{n-1}(a)\right]$
tend toward 1 if and only if $a\to+\infty$.

 \bibliographystyle{apsrev}
\bibliography{ref}

\end{document}